\documentclass[fleqn,usenatbib]{mnras}

\usepackage{newtxtext,newtxmath}
\usepackage{graphicx}
\usepackage{bm}
\usepackage{cancel}
\usepackage{epsfig}
\usepackage{dcolumn}
\usepackage{amsmath}
\usepackage{hhline}
\usepackage{enumerate}
\usepackage[utf8]{inputenc}
\usepackage[dvipsnames]{xcolor}
\usepackage{hyperref}
\usepackage{wasysym}
\usepackage{tikz} 
\usepackage[T1]{fontenc}

\DeclareRobustCommand{\VAN}[3]{#2}
\let\VANthebibliography\thebibliography
\def\thebibliography{\DeclareRobustCommand{\VAN}[3]{##3}\VANthebibliography}

\def \beq  {\begin{equation}}
\def \eeq  {\end{equation}}
\def \ber  {\begin{eqnarray}}
\def \eer  {\end{eqnarray}}

\def \om    {\Omega}

\def \om0m {\Omega_{0\rm m}}

\title{Late time approaches to the Hubble tension deforming $H(z)$, worsen the growth tension}

\author[G. Alestas \& L. Perivolaropoulos]{
George Alestas \orcidA{},$^{1}$\thanks{Contact e-mail: \href{mailto:g.alestas@uoi.gr}{g.alestas@uoi.gr}}%
Leandros Perivolaropoulos \orcidA{},$^{1}$\thanks{Contact e-mail: \href{mailto:leandros@uoi.gr}{leandros@uoi.gr}}
\\
$^{1}$Department of Physics, University of Ioannina, GR-45110, Ioannina, Greece
}

\pubyear{2021}

\begin{document}

\interfootnotelinepenalty=10000

\newcommand{\newc}{\newcommand}

\newcommand{\orcidauthorA}{0000-0003-1790-4914} 
\newcommand{\orcidauthorB}{0000-0001-9330-2371} 

\newcommand{\be}{\begin{equation}}
\newcommand{\ee}{\end{equation}}
\newcommand{\ba}{\begin{eqnarray}}
\newcommand{\ea}{\end{eqnarray}}
\newcommand{\bea}{\begin{eqnarray*}}
\newcommand{\eea}{\end{eqnarray*}}
\newc{\D}{\partial}
\newc{\ie}{{\it i.e.} }
\newc{\eg}{{\it e.g.} }
\newc{\etc}{{\it etc.} }
\newc{\etal}{{\it et al.}}
\newc{\lcdm}{$\Lambda$CDM }
\newc{\lcdmnospace}{$\Lambda$CDM}
\newc{\wcdm}{$w$CDM }
\newc{\plcdm}{Planck18/$\Lambda$CDM }
\newc{\plcdmnospace}{Planck18/$\Lambda$CDM}
\newc{\omom}{$\Omega_{0m}$ }
\newc{\omomnospace}{$\Omega_{0m}$}
\newcommand{\nn}{\nonumber}
\newc{\ra}{\Rightarrow}
\newc{\baodv}{$\frac{D_V}{r_s}$ }
\newc{\baodvnospace}{$\frac{D_V}{r_s}$}
\newc{\baoda}{$\frac{D_A}{r_s}$ } 
\newc{\baodanospace}{$\frac{D_A}{r_s}$}
\newc{\baodh}{$\frac{D_H}{r_s}$ }
\newc{\baodhnospace}{$\frac{D_H}{r_s}$}

\newcommand{\orcidicon}{\includegraphics[width=0.32cm]{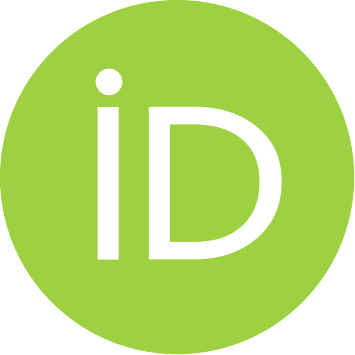}}

\foreach \x in {A, ..., Z}{%
\expandafter\xdef\csname orcid\x\endcsname{\noexpand\href{https://orcid.org/\csname orcidauthor\x\endcsname}{\noexpand\orcidicon}}
}

\label{firstpage}
\pagerange{\pageref{firstpage}--\pageref{lastpage}}
\maketitle

\begin{abstract}
Many late time approaches for the solution of the Hubble tension use late time smooth deformations of the Hubble expansion rate $H(z)$ of the Planck18/$\Lambda$CDM best fit to match the locally measured value of $H_0$ while effectively keeping the comoving distance to the last scattering surface and $\Omega_{0m} h^2$ fixed to maintain consistency with Planck CMB measurements. A well known problem of these approaches is that they worsen the fit to low $z$ distance probes. Here we show that another problem of these approaches is that they worsen the level of the $\Omega_{0m}-\sigma_8$ growth tension. We use the generic class of CPL parametrizations corresponding to evolving dark energy equation of state parameter $w(z)=w_0+w_1\frac{z}{1+z}$ with local measurements $H_0$ prior and identify the pairs $(w_0, w_1)$ that satisfy this condition. This is a generic class of smooth deformations of $H(z)$ that are designed to address the Hubble tension. We show that for these models the growth tension between dynamical probe data and CMB constraints is worse than the corresponding tension of the standard Planck18/$\Lambda$CDM model. We justify this feature using a full numerical solution of the growth equation and fit to the data, as well as by using an approximate analytic approach. The problem does not affect recent proposed solutions of the Hubble crisis involving a SnIa intrinsic luminosity transition at $z_t\simeq 0.01$.    
\end{abstract}




\section{Introduction}

The expansion rate of the universe $H(z)$ at redshift $z\in [0.01,1100]$ has been measured using locally calibrated supernovae Ia as standard candles and using the sound horizon at recombination as a standard ruler calibrated by the peak locations of the CMB anisotropy spectrum. The two approaches have been extensively tested, appear to be robust and free from major systematics and agree on a shape of $H(z)$ that is consistent with the \plcdm form
\be 
H(z)^2=H_0^2\left[\Omega_{0m}(1+z)^3 + (1-\Omega_{0m})\right]
\label{lcdm}
\ee
with a matter density parameter $\Omega_{0m}=0.315\pm 0.007$. However, the scale $H(z=0)=H_0$ (the {\it Hubble constant}) obtained by the sound horizon approach ($H_0^{P18}=67.4 \pm 0.5 \;km\; sec^{-1}\; Mpc^{-1}$ \citep{Aghanim:2018eyx}) is lower by $9\%$ compared to the Hubble constant obtained by the SnIa distance ladder approach ($H_0^{R20}=74.03 \pm 1.42 \;km\; sec^{-1}\; Mpc^{-1}$ \citep{Riess:2019cxk}) assuming that the absolute luminosity of SnIa remains unchanged before and after $z=0.01$\footnote{This is a crucial assumption that has been put under intense scrutiny recently \citep{Alestas:2020zol,Marra:2021fvf}}. This is a statistically significant inconsistency at more that $4\sigma$ level and constitutes the most important problem of modern cosmology.

A significant number of theoretical approaches attempting to address this issue have been proposed \citep{DiValentino:2021izs, Kazantzidis:2019dvk}, and can be divided in three broad classes:
\begin{itemize}
\item
'Early time' models that attempt to recalibrate the scale of the standard ruler (the sound horizon at recombination) by introducing new physics during the prerecombination epoch that deform $H(z)$ at prerecombination redshifts $z>1100$ (early dark energy \citep{Karwal:2016vyq,Poulin:2018cxd}, new types of neutrinos \citep{Sakstein:2019fmf} etc). The challenge for this class of models is that they tend to predict stronger growth of perturbations than implied by dynamical probes like redshift space distortion (RSD) and weak lensing (WL) data and thus they worsen the so called '$\Omega_{0m}-\sigma_8$ growth tension \citep{Jedamzik:2020zmd} (even though this issue is still under debate \citep{Smith:2020rxx}). This tension emerges by the observational fact that dynamical cosmological probes favor weaker growth of perturbations than geometric probes in the context of general relativity and the \plcdm standard model \citep{Hildebrandt:2016iqg,Nesseris:2017vor,Macaulay:2013swa,Kazantzidis:2018rnb,Skara:2019usd, Kazantzidis:2019dvk, Perivolaropoulos:2019vkb}.
\item
Late time deformations of the expansion rate $H(z)$ that attempt to deform the \plcdm $H(z)$ at late times, so that it keeps its consistency with the CMB anisotropy spectrum while ending at the locally measured value of $H(z=0)=H_0^{R20}$. The challenge for this class of models is that for smooth $H(z)$ deformations they have difficulty fitting low $z$ cosmological distance measurements obtained by BAO and SnIa data and thus they also can not fully resolve the Hubble problem \citep{Benevento:2020fev,Alestas:2020mvb,Yang:2021flj}.
\item
Late time transitions at a redshift $z_t\simeq 0.01$ of the SnIa absolute magnitude $M$ to a lower value (brighter SnIa at $z>z_t$) by $\Delta M\simeq -0.2$ have also been recently proposed as an approach to the Hubble problem \citep{Alestas:2020zol, Marra:2021fvf}. Such a reduction of $M$ at $z>z_t$ may be induced \eg by a corresponding transition of the effective gravitational constant $G_{eff}$ leading to an increase of the SnIa intrinsic luminosity at $z>z_t$ \citep{Marra:2021fvf}. This type of transition\footnote{Hints of a possible evolution of the absolute magnitude M have been recently identified in Refs. \citep{Kazantzidis:2020tko, Kazantzidis:2020xta, Dainotti:2021pqg}} could coexist with a transition of the dark energy equation of state $w$ from $w=-1$ at $z>z_t$ to a value less than $-1$ at $z<z_t$ (phantom transition) which if present, would allow for a lower magnitude $\Delta M$ of the $M$ transition. This class of models could fully resolve the Hubble problem while at the same time address the growth tension by reducing the growth rate of cosmological perturbation due to the lower value of $G_{eff}$ at $z>z_t$ \citep{Marra:2021fvf}. This class of models is highly predictive and may by challenged by upcoming and existing cosmological and astrophysical data (\eg standard sirens and Tully-Fisher data). A challenge for this class of models is also the identification of observationally viable theoretical models that can support this transition.
\end{itemize}

In the present analysis we focus on late time smooth deformation of $H(z)$ models that address the Hubble tension and discuss the following question: {\it 'Can this class of models improve the growth tension by decreasing the growth rate of cosmological perturbations compared to the \plcdm model?' }

In order to address this question we consider a generic class of $H(z)$ deformation models based on a first order expansion of the dark energy equation of state $w$ around the present value of the scale factor $a=1$,  $w(a)=w_0+w_1(1-a)$ known as the CPL \citep{Chevallier:2000qy,Linder:2002et} parametrization which in redshift space is expressed as
\be
w(z)=w_0 +w_1 \frac{z}{1+z}
\label{cpl}
\ee
We impose consistency with the Planck anisotropy spectrum and local measurements of $H(z)$ by three conditions \citep{Alestas:2020mvb}:  Firstly, fixing $\omega_m\equiv \Omega_{0m}h^2$ ($h\equiv H_0/100km\, s^{-1}\, Mpc^{-1}$) to the \plcdm value $ \omega_m={\bar \omega}_m\equiv 0.143$. Secondly, fixing the comoving distance to recombination (flat space) $r(z_{rec})\equiv \int_0^{z_{rec}} \frac{dz}{H(z}$ to its \plcdm value. Thirdly, fixing the value of $H_0$ to its locally measured value $H_0^{R20}$. 

These conditions lead to the numerical evaluation of the function $w_{1h}(w_0)$ such that for any given value of $w_0$ we obtain the corresponding value of $w_1=w_{1h}$ that can potentially address the Hubble problem by fitting local measurements of $H_0$ while being consistent with the CMB anisotropy spectrum \citep{Alestas:2020mvb}. We then focus on pairs $(w_0,w_{1h}(w_0))$, evaluate the predicted growth factor of perturbations in the context of general relativity $\frac{\delta(z=0)}{\delta(z_{rec})}$ and compare it with the corresponding growth factor predicted by the best fit \plcdm $H(z)$. We thus address the question: {\it Are there $w_0,w_1$ pairs that can potentially address the Hubble problem while having lower predicted growth of perturbations than the \plcdm form of $H(z)$ which is already in tension with RSD and weak lensing data? }

In the special case of $w_{1h}=0$ corresponding to $w_0\simeq -1.22$ ($wCDM$), this question was addressed in \citep{Alestas:2020mvb} where it was shown that the growth tension increases in this $wCDM$ model with parameter values chosen in a way to address the Hubble tension. Here, we generalize that analysis to more general smooth deformations of $H(z)$.

In addition to evaluating the growth factor for various $w_0-w_1$ parameter values that address the Hubble tension, we use a robust RSD $f\sigma_8$ data compilation \citep{Nesseris:2017vor,Sagredo:2018ahx} to construct the $\sigma_8-\Omega_{0m}$ likelihood contours for representative $(w_0,w_{1h}(w_0))$ pairs to identify the tension level with the corresponding Planck likelihood contours and find how does the tension change as we move in the parameter space $(w_0-w_{1h}(w_0))$ that can potentially address the Hubble problem. Notice that this parameter space, and all similar smooth $H(z)$ deformations, can make local measurements of $H_0$ consistent with the CMB spectrum but do not fit well the low $z$ distance data (BAO and SnIa) as it has been demonstrated in previous studies \citep{Alestas:2020mvb, DiValentino:2020naf, DiValentino:2017zyq}. We confirm these results by fitting these models to SnIa, BAO and CMB data and demonstrating that the fit is significantly worse than the corresponding fit of the standard \plcdm model. This well known problem of this class of models with respect to the BAO and SnIa data is independent from the main issue pointed out in the present analysis which refers to the consistency of this class of models with the growth $f\sigma_8$ data.

Finally, we use the Pantheon SnIa compilation to identify the best fit value of the SnIa absolute magnitude $M$ in the context of these $H(z)$ deformations that address the Hubble tension. We compare these best fit values with the corresponding range of $M$ implied by Cepheid calibrators and search for a possible new type of problem for these models (the $M$ tension \citep{Camarena:2021jlr, Camarena:2019moy}).

\begin{figure*}
\centering
\includegraphics[width = 1\textwidth]{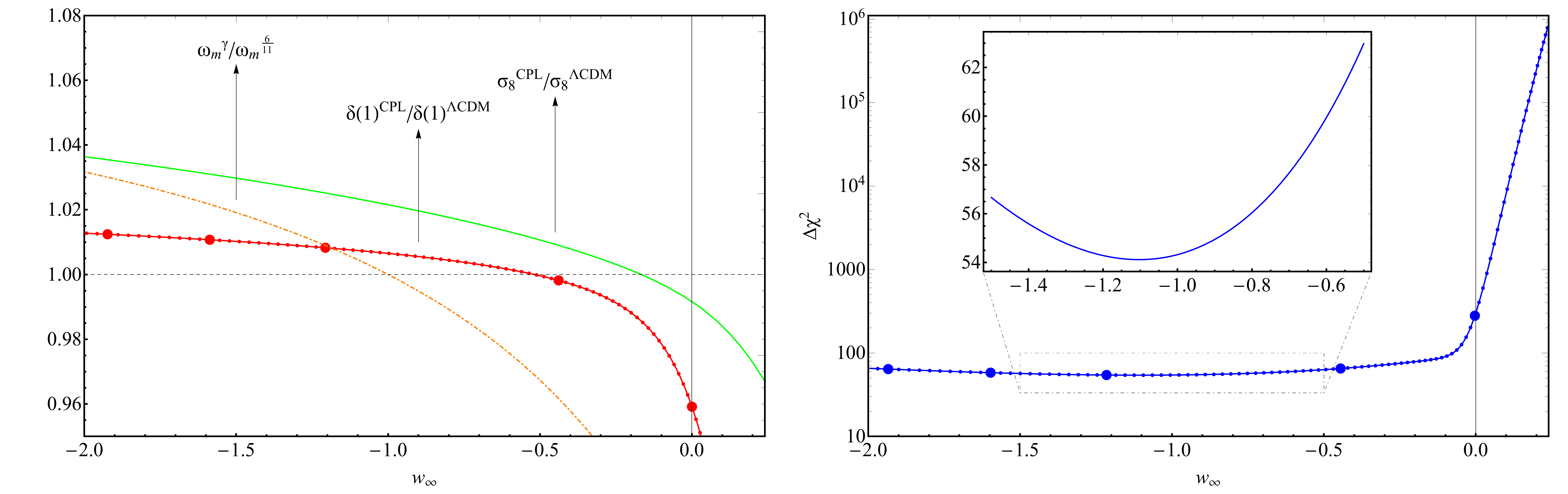}
\caption{Left panel:The relative growth factor $\delta(1)^{CPL}/\delta(1)^{\Lambda CDM}$ (red line), the best fit ratio $\sigma^{CPL}_{8}/\sigma^{\Lambda CDM}_{8}$ (green line, fixed $\omega_m=0.143$) and the ratio $\omega_{m}^{\gamma}/\omega_{m}^{\frac{6}{11}}$ (orange dot-dashed line) all have similar dependence on $w_{\infty}$. The five thick dots correspond to the CPL parameter values $(w_0, w_1)$ pairs, $(-1, -0.93)$, $(-1.1, -0.497)$, $(-1.22, 0)$, $(-1.5, 1.05)$ and $(-1.73, 1.72)$ of the five panels of Fig. \ref{fig3} (the sixth corresponds to \lcdm).
Right panel:
The quality of fit to the CMB shift parameters, Pantheon and BAO data compared to \lcdm is significantly worse than \lcdm for the CPL models that address the Hubble tension. In all cases $\Delta \chi^2 >50$. The minimum ($\Delta \chi^2 \simeq 54$) occurs at about $w_{\infty}\simeq -1.2$ corresponding to $wCDM$.
}
\label{fig1}
\end{figure*} 

\section{Hubble tension and the CPL parametrization}

A deformation of the Hubble expansion rate from its \plcdm form (\ref{lcdm}) may be expressed as
\be
H(z,\omega_m,\omega_r,h,w(z))=H_0 \sqrt{\begin{aligned}
    &\Omega_{0m} (1+z)^3 + \Omega_{0r} (1+z)^4 \\ + &\Omega_{0 de} e^{3\int_0^z dz'\; (1+w(z'))/(1+z')}\end{aligned}}
\label{hz}
\ee
where $w(z)$ is the dark energy equation of state parameter at redshift $z$, $\Omega_{0r}$, $\Omega_{0m}$ are the present day radiation and matter density parameters and $\Omega_{0de}=1-\Omega_{0m}-\Omega_{0r}$ is the present day value of the dark energy density parameter assuming spatial flatness. We also define $\omega_r\equiv \Omega_{0r} h^2 $. This deformed Hubble expansion can become simultaneously consistent with local measurements of $H_0=H_0^{R20}$ as well as with the CMB anisotropy spectrum, provided that the following conditions are satisfied \citep{Efstathiou:1998xx,Elgaroy:2007bv,Alestas:2020mvb} 
\begin{itemize}
\item The matter and radiation density parameter combinations $\omega_{m}$ and $\omega_{r}$ are fixed to their \plcdm best fit value $\bar \omega_m = 0.1430 \pm 0.0011$ and $\bar\omega_r = (4.64 \pm 0.3)\; 10^{-5}$.
\item
The cosmological comoving distance to the recombination redshift
\be
r(z_r,\omega_m,\omega_r,w(z))=\int_0^{z_{r}}\frac{dz}{H(z)}=\int_{a_r}^1\frac{da'}{a'^2 H(a')}
\label{comd}
\ee
($a$ is the cosmic scale factor, $z_r\simeq 1091$ is the redshift of recombination) is fixed to the \plcdm best fit value $\bar r = (100 \; km \; sec^{-1} \; Mpc^{-1})^{-1} (4.62\pm 0.08)$.
\item
The Hubble parameter $H_0$ is fixed to its locally measured value $H_0^{R20}$.
\end{itemize}
These conditions correspond to a constraint on the dark energy equation of state $w(z)$. For example in the context of the typical deformation model corresponding to the CPL parametrization (\ref{cpl}) where $H(z)$ is of the form
\be
H(z)=H_0 \sqrt{\begin{aligned}\om0m &(1+z)^3 +\Omega_{0 \rm r} (1+z)^4 \\ & +\left(1-\om0m - \Omega_{0 \rm r}\right) 
(1+z)^{3(1+w_0+w_1)}  e^{-3 \frac{w_1 z}{1+z}}\end{aligned}} \label{hzcpl}
 \ee
this constraint may be approximately expressed as \citep{Alestas:2020mvb}
\be
w_{1h}(w_0)\simeq -4.17 w_0 -5.08
\label{w1ht}
\ee
which defines a set of points in the CPL parameter space which can in principle address the Hubble tension by a deformation of $H(z)$. More accurate values of the $w_1(w_0)$ dependence may be obtained by numerical solution of the equation $r(z_{r},{\bar \omega}_m,{\bar \omega}_r,w(z))=4.62$. In what follows we refer to eq. (\ref{w1ht}) but we actually use these more accurate numerically obtained values to identify the $w_0-w_1$ pairs that can potentially address the Hubble tension.

In practice, a challenge faced by this late time deformation approach is the relatively poor fit it provides to local distance measurements at $z<2$ by BAO \citep{Beutler:2011hx,Ross:2014qpa,Alam:2016hwk} and SnIa data \citep{Scolnic:2017caz}.  The fit to these local measurement data becomes dramatically worse when the asymptotic value of $w(z)$ at early times which in the CPL case is 
\be
w_\infty = w_0+w_1
\label{winfty}
\ee
increases to values $w_\infty>-0.5$ while the best possible fit to BAO-SnIa is obtained for $w_\infty\simeq -1.2$ (even in this case however, the fit quality is significantly worse than \plcdm). 

\begin{figure*}
\centering
\includegraphics[width = 0.9 \textwidth]{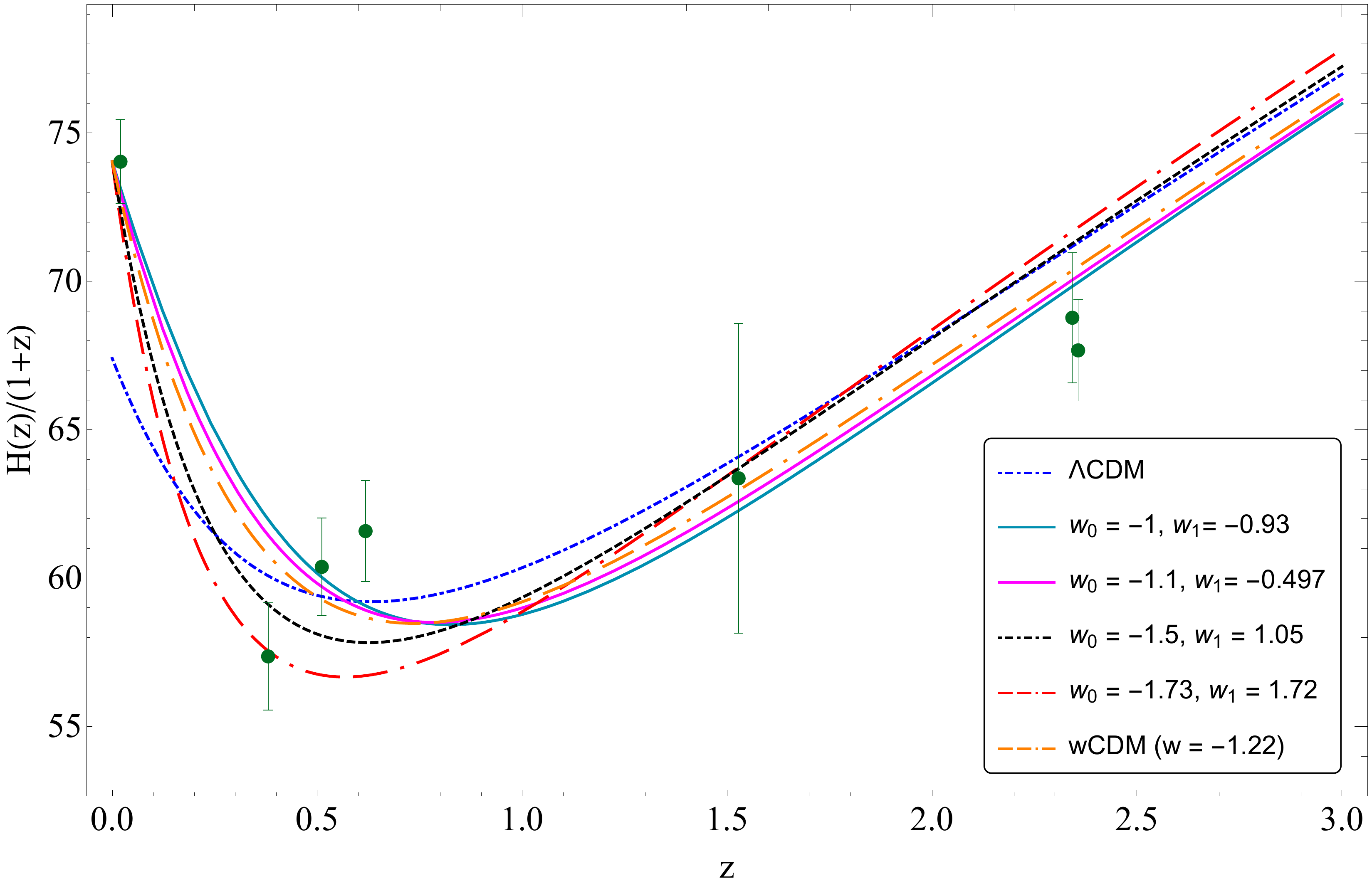}
\caption{The \plcdm form of $H(z)/(1+z)$ (blue dot-dashed line) is compared with the same function obtained with various pairs of CPL parameters that address the Hubble tension. Some BAO data are also shown.}
\label{fig2}
\end{figure*}

In the next section we assume the validity of eq. (\ref{w1ht}) and investigate the  growth of perturbations and the level of the growth tension, in the context of this set of $H(z)$ deformations that can potentially address the Hubble tension. In particular we compare the growth factor of these models with the corresponding growth factor of \plcdm and identify the tension level in the  $\Omega_{0m}-\sigma_8$ parameter space between RSD growth data and \plcdm likelihoods contours.

\section{Growth of perturbations and the growth tension in Hubble deformation models}
\subsection{Evolution of matter density perturbations}
The evolution of the growth factor of cosmological matter perturbations $\delta(a)\equiv \frac{\delta \rho}{\rho}(a)$ in terms of the cosmic scale factor $a$ is determined in subhorizon scales by  the following equation \citep{DeFelice:2010gb, Tsujikawa:2007gd, DeFelice:2010aj, Nesseris:2009jf, Nesseris:2017vor, Linder:2003dr}:
\be
\delta''(a)+\left(\frac{3}{a}+\frac{H'(a)}{H(a)}\right)\delta'(a)
-\frac{3\Omega_{0m}}{2a^5 H(a)^2/H_0^2}~\delta(a)=0,
\label{eq:ode}
\ee
where primes denote differentiation with respect to the scale factor $a$ and $H(a)\equiv\frac{\dot{a}}{a}$ is the Hubble expansion rate. The initial conditions for the solution of eq. (\ref{eq:ode}) are usually taken deep in the matter era (\eg for $a_i=0.001$) where it is easy to show that $\delta(a_i)\sim a_i$. The growth factor $\delta(a)/\delta(a_i)$ indicated by this equation in the context of \plcdm best fit parameters is higher than the growth favored by dynamical probe data like weak lensing \citep{Schmidt:2008hc,kids1,cfhtlens,Joudaki:2017zdt,Troxel:2017xyo,kids2,des3,Abbott:2018xao}, cluster counts \citep{Rozo:2009jj,Rapetti:2008rm,Bocquet:2014lmj,Ruiz:2014hma} and redshift space distortions \citep{Samushia:2012iq,Macaulay:2013swa,Johnson:2015aaa,Nesseris:2017vor,Kazantzidis:2018rnb} at a $2-3\sigma$ level. This is known as the growth tension or $\Omega_{0m}-\sigma_8$ tension where $\sigma_8$ is defined as the matter density rms fluctuations within spheres of radius $8 h^{-1}$Mpc at the present time $z=0$ and is connected with the amplitude of the primordial fluctuation spectrum. In particular the best fit value of the matter density parameter favored by \plcdm is higher than the value favored by the dynamical probes. This indicates that dynamical probes prefer a weaker growth of perturbations since the matter density parameter effectively 'drives' the growth of density perturbations.

A useful bias-free statistic probed by RSD data is the product $f\sigma_8$:
\begin{align}
f\sigma_8(a)=\frac{\sigma_8}{\delta(a=1)}~a~\delta'(a,\Omega_{0m}) \,,
\label{eq:fs8}
\end{align}
where $f\equiv \frac{d\ln \delta}{d\ln\,a}$ is the growth rate of matter density perturbations.
Notice that for a given measured value of $f\sigma_8(a)$, weaker growth (smaller $\delta(a=1)$) implies a lower value of $\sigma_8$ (assuming that $\delta'(a)$ does not change significantly for a given value of $a$). This is demonstrated in Fig. \ref{fig1} where we show the properly normalized best fit value of $\sigma_8$ (green line) obtained by fitting the solution of eq. (\ref{eq:ode}) to the robust $f\sigma_8$ dataset of Ref. \citep{Nesseris:2017vor} \footnote{This dataset is optimized for independence of datapoints but it involves significanty less datapoints than the more complete compilation of Ref. \citep{Skara:2019usd}.} for various values of $w_\infty$ under the assumption of $\omega_m=0.143$, $h=0.74$ and eq. (\ref{w1ht}),  conditions required for consistency of local measurements of $H_0$ and \plcdm anisotropy spectrum.  Clearly $\sigma_8(w_\infty)/\sigma_8^{P18\Lambda}$ ($\sigma_8^{P18\Lambda}$ denotes the best fit value in the context of \plcdm) has the same monotonicity and differs by less than $2\%$ from $\delta(1,w_\infty)/\delta(1)^{P18\Lambda}$ (red line) thus justifying that the best fit $\sigma_8$ and $\delta(a=1)$ are approximately proportional.

\subsection{Analytic approximate solutions}
An approximate solution to eq. (\ref{eq:ode}) can be found \citep{Linder:2007hg} by utilizing a growth index $\gamma$, which is used to parameterize the linear growing mode of models with time varying equations of state, such as eq. (\ref{cpl}). Using $\gamma$ and ignoring the effects of radiation, the growth factor solution $\Delta(a)\equiv \frac{\delta(a=1)}{\delta(a_i)}$ of (\ref{eq:ode}) may be approximated as \citep{Linder:2007hg,Basilakos:2008tk}
\be
\Delta(a)={\rm exp} \left[\int_{a_i}^{a} \frac{\Omega_{m}^{\gamma}(a')}{a'} {d}a' \right]
\label{glind} 
\ee
where $\Delta(a)$ is the normalized growth factor $\delta(a)/\delta(a_i)$, $a_i=0.001$ is an initial redshift deep in the matter era when $\delta(a)\sim a$ and
\be
\Omega_m(a)\equiv \frac{\Omega_{0m} H_0^2 a^{-3}}{H(a)^2}=\frac{\omega_m \, a^{-3}}{h(a)^2}
\label{oma}
\ee
with $h(a)^2\equiv \omega_m a^{-3} +(h^2-\omega_m)f_a(a)$ ($f_a(a)$ denotes the evolution of the dark energy density).
The growth index is approximated by \citep{Linder:2007hg}
\be
\gamma=\frac{6-3(1+w_{\infty})}{11-6(1+w_{\infty})}. 
\ee 
where $w_{\infty}$ is defined in eq. (\ref{winfty}). For \lcdm ($w=-1$) we have $\gamma=6/11\simeq 0.55$. From eq. (\ref{glind}) it is easy to obtain the well known approximate expression for the growth rate $f(a)$ of density perturbations
\be
f(a)\equiv \frac{d\ln \Delta}{d\ln a}\simeq \Omega_m(a)^\gamma
\label{fgamma}
\ee
which may also be used as a definition of the growth index $\gamma$.

\begin{figure*}
\centering
\includegraphics[width = 1 \textwidth]{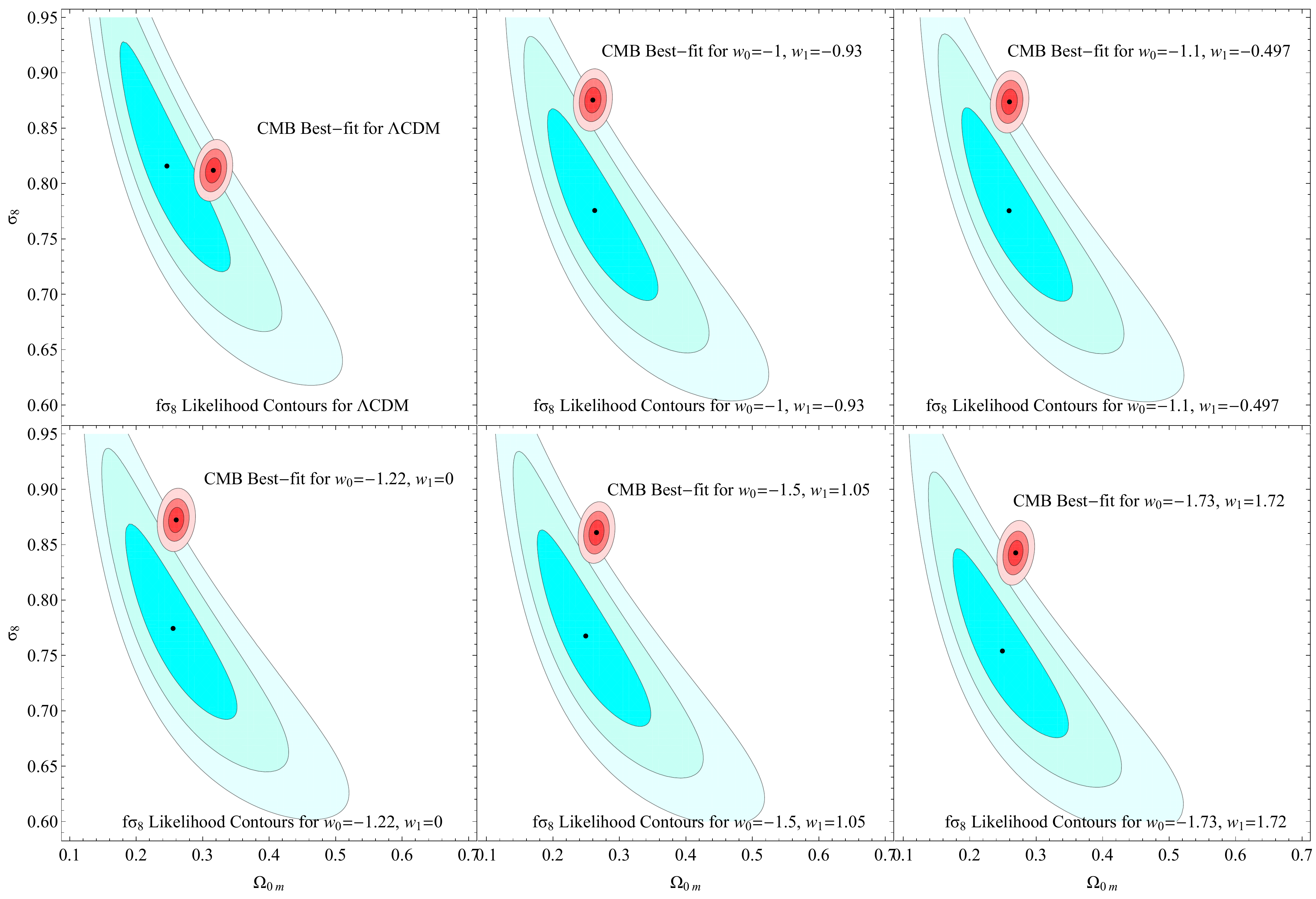}
\caption{The cyan and the red contours correspond to the Growth and the Plank 18 CMB data respectively, for the \lcdm and various $(w_0, w_1)$ pairs of the CPL model. The $\Delta\sigma$ differences between the best fit values produced by the two contours in each case, are shown in Tab. \ref{tab:deltasigmaM}. It is clear that the tension does not ease in the case of the CPL model despite the fact that it appears to solve the $H_0$ tension for the same $w_0$ and $w_1$ values used.}
\label{fig3}
\end{figure*} 

Using eqs. (\ref{glind}), (\ref{oma}) it is easy to express the growth factor as
\be
\Delta(a)=\frac{\delta(a)}{\delta(a_i)}={\rm exp} \left[\omega_m^\gamma\,\int_{a_i}^{a} \frac{da'}{a'^{1+3\gamma}\,h(a')^{2\gamma}} \right]
\label{glind1} 
\ee
Since $\gamma\in[0.45,0.65]$ in most physically interesting cases, the integral in the exponential of eq. (\ref{glind1}) is very similar to the integral of the comoving distance (\ref{comd}) \citep{Basilakos:2008tk}. Since the dark energy parameter values (\eg pairs of $w_0-w_1$ in the CPL case) that can address the Hubble tension have approximately fixed comoving distance to recombination they should also have approximately fixed growth integral in eq. (\ref{glind1}) for $a=1$. Therefore, the growth factor $\Delta(a=1)$ is expected to have approximately similar behavior as $\omega_m^{\gamma(w_\infty)}$. This is demonstrated in Fig. \ref{fig1} where we show the growth factor $\delta(1)^{CPL}/\delta(1)^{P18\Lambda}$ as obtained by a numerical solution of eq. (\ref{eq:ode}) using the \plcdm best fit parameter values ($\delta(1)^{P18\Lambda}$) and the CPL parameter values (\ref{w1ht}) that address the Hubble problem ($\delta(1)^{CPL}$). In both case we fixed $\omega_m=0.143$ for consistency with the CMB anisotropy spectrum while we set $h=0.74$ for $\delta(1)^{CPL}$ and $h=0.67$ for $\delta(1)^{P18\Lambda}$ in eq. (\ref{eq:ode}). Superimposed is the ratio $\omega_m^{\gamma(w_\infty)}/\omega_m^{\gamma(w_\infty=-1)}$ for $\omega_m=0.143$ and $\gamma(w_\infty=-1)=\frac{6}{11}$ corresponding to the \lcdm growth index. The two quantities ($\omega_m^{\gamma(w_\infty)}/\omega_m^{\gamma(w_\infty=-1)}$ and $\delta(1)^{CPL}/\delta(1)^{P18\Lambda}$) have similar monotonicities and  differ by less than $4\%$ in the range $w_\infty \in [-2,-0.5]$. This validates the approximation that the growth integral of eq. (\ref{glind1}) varies slowly with $w_\infty$ when (\ref{w1ht}) is obeyed.

As shown in Fig. \ref{fig1} (red curve) the $H(z)$ CPL deformations that can address the Hubble tension induce a growth factor that is larger than the one implied by a \plcdm background for all $w_\infty<-0.5$. This range of $w_\infty$ includes all the values of parameters which are consistent with SnIa, BAO and CMB data. This is demonstrated in Fig. \ref{fig1} (right panel) where we show the excess value of $\chi^2$ with respect to \plcdm as a function of $w_\infty$ using the Pantheon SnIa data along with a compilation of 9 BAO datapoints and two CMB effective distance/shift parameters \citep{Alestas:2020zol}. Clearly, the best fit is obtained for $w_\infty\simeq -1.1\pm 0.2$, while the value $w_\infty=-0.5$ is more than $3\sigma$ away from the best fit value. 

The strong deformation of $H(z)$ implied by models with high values of $w_\infty$ is also shown in Fig. \ref{fig2} where the \plcdm form of $H(z)/(1+z)$ (blue dot-dashed line) is compared with the same function obtained with various pairs of CPL parameters that address the Hubble tension. Clearly the strongest deformation at low $z$ occurs for models with $w_\infty<-0.5$ which implies also inconsistency with BAO and SnIa data at redshifts of $O(1)$.

\subsection{Fit to $f\sigma_8$ data: Increased tension with \plcdm}

\begin{centering}
\begin{table}
\caption{\label{tab:deltasigmaM}
The three problems of $H(z)$ deformations addressing the Hubble crisis (columns 1,2). Column 3: The deviation of the best fit value of the absolute magnitude $M$ for each deformation, from the Cepheid calibrated value of \citep{Camarena:2019moy, Camarena:2021jlr} shown in Fig. \ref{fig4}. Column 4: The $\Delta\chi^{2}$ differences with respect to Planck18/$\Lambda$CDM shown also in Fig. \ref{fig1} (right panel) for each $(w_0, w_1)$ pair that addresses the Hubble tension. Column 5: The $\Delta\sigma$ differences between the best fit values of the Growth and CMB data contours depicted in Fig. \ref{fig3}.}
\begin{tabular}{|c|c|c|c|c|}
\hline 
 \rule{0pt}{3ex}  
$w_0$ & $w_1$ & $\Delta M$ & $\Delta\chi^{2}$ & $\Delta\sigma$ \\
    \hline
    \rule{0pt}{3ex}  
 -1 & 0 & -0.19 & - & 2 \\
 -1 & -0.93 & -0.02 & 63 & 2.9 \\
 -1.1 & -0.50 & -0.03 & 57 & 3.0 \\
 -1.22 & 0.0 & -0.05 & 54 & 3.1 \\
 -1.50 & 1.05 & -0.09 & 65 & 3.4 \\
 -1.73 & 1.72 & -0.12 & 279 & 3.4 \\
 \hline
 \end{tabular}
\end{table}
\end{centering}

The increased tension level between CMB data and RSD growth data in the context of late time $H(z)$ deformations addressing the Hubble tension is demonstrated in Fig. \ref{fig3} where we show the CMB data likelihood contours (Planck18 chains) in the parameter space $\Omega_{0m}-\sigma_8$ superimposed with the corresponding contours obtained from a robust compilation of RSD $f\sigma_8$ data \citep{Nesseris:2017vor,Sagredo:2018ahx} for  \lcdm (upper left pannel), and five  CPL $w_0,w_1$ parameter pairs that can address the Hubble problem with $h=0.74$. These five pairs (thick dots in Fig. \ref{fig1}) in addition to being disfavored by low z geometric probes (BAO and SnIa) by $\delta \chi^2>50$  (see Tab. \ref{tab:deltasigmaM}), also lead to increased tension between CMB and growth data compared to \lcdm as shown in Fig. \ref{fig3} and Tab. \ref{tab:deltasigmaM} even for parameter values where the growth factor is less than that of \plcdm (lower right panel corresponding to $w_\infty<-0.5$). In constructing Fig. \ref{fig3} and Tab. \ref{tab:deltasigmaM} we have only fixed the parameters $w_0$, $w_1$ in each panel as indicated so that the Hubble tension is addressed (in the 5 panels) but have left free $\Omega_{0m}$ and $\sigma_8$ to be fitted by the data. Notice that in all panels the CMB data favor a value of $\omega_m\simeq 0.143$ as expected.

\begin{figure}
\centering
\includegraphics[width = 0.45 \textwidth]{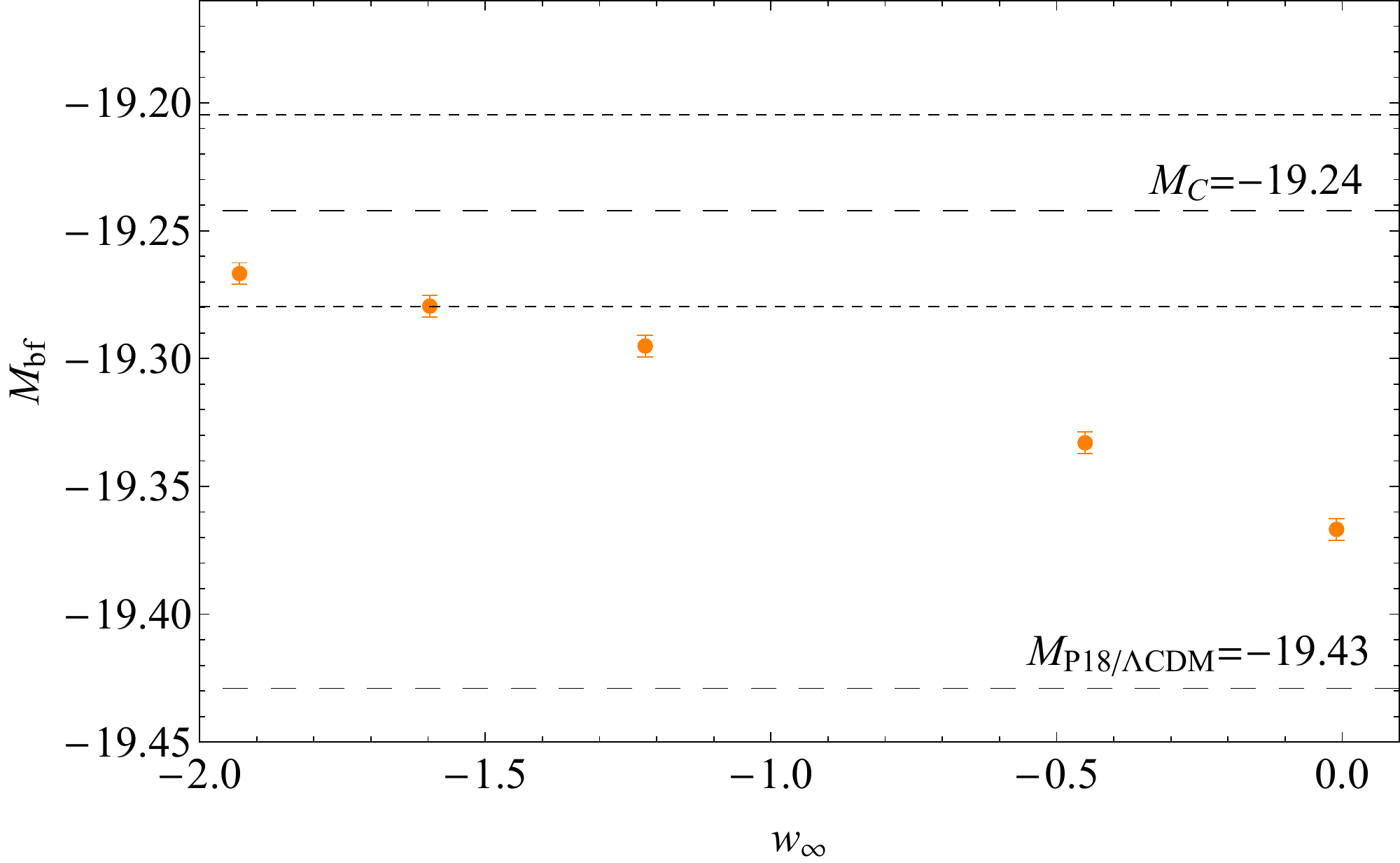}
\caption{The best fit values of the absolute magnitude $M$, for the $(w_0, w1)$ pairs displayed in Tab. \ref{tab:deltasigmaM}. These values are consistently lower than the corresponding value of $M$ implied by local Cepheid calibrators (upper dashed line) even though this tension is not as large as for the best fit value of $M$ obtained in the context of the standard \plcdm model (lower dashed line).} 
\label{fig4}
\end{figure} 

In addition to the reduced quality of fit to low $z$ geometric probes and the increased growth tension, the $H(z)$ deformation models addressing the Hubble tension face another challenge: They lead to a best fit value of the SnIa absolute magnitude $M$ that is consistently lower than the corresponding value implied by the Cepheid calibrators at $z<0.01$ $M=-19.24\pm 0.04$. This difference is indicated in Fig. \ref{fig4} and in Tab. \ref{tab:deltasigmaM}. 

\begin{figure}
\centering
\includegraphics[width = 0.45 \textwidth]{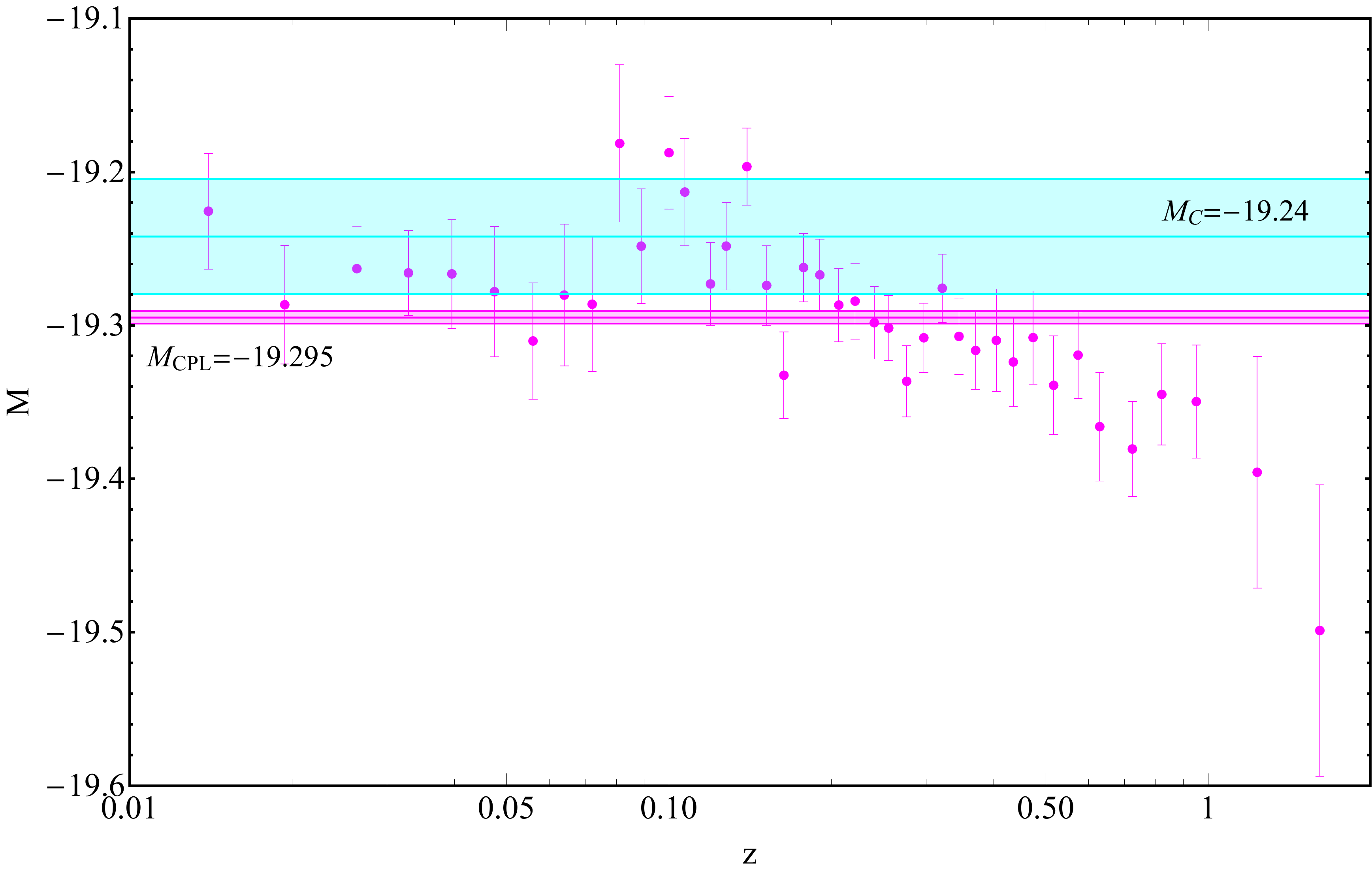}
\caption{The values of the absolute magnitude corresponding to the binned SnIa data in the context of the CPL model $(w_0=-1.22,w_1=0)$ that addresses the $H_0$ tension via an $H_0$ prior, shown to be in tension with the Cepheid calibrated $M$ range.} 
\label{fig5}
\end{figure} 

\section{Conclusions}

We have demonstrated that late time deformations of $H(z)$ designed to address the Hubble tension not only worsen the fit to low z geometric probes like SnIa and BAO data but also worsen the tension between CMB and growth data in he $\Omega_{0m}-\sigma_8$ parameter space. A similar effect occurs for early time approaches to the Hubble problem, which may also worsen the growth tension \citep{Jedamzik:2020zmd}. In addition to these problems we have shown that these models lead to lower best fit values of the SnIa absolute magnitude than the Cepheid calibrator absolute magnitude.


One could argue that there are several physical cases not covered by the CPL model \citep{Scherrer:2015tra, Busti:2015xqa}, thus casting a doubt on the generality of our results. However these issues arise due to CPL being accurate only around the present day value of the scale factor, since it is a linear expansion on $w$ around the present. The fact that the CPL parametrization is ultimately a good qualitative representative of smooth $H(z)$ deformations that address the Hubble tension \citep{Yang:2021flj}, can also be seen in Fig. \ref{fig2} where it is obvious that the parameters that solve the aforementioned tension correspond to a wide range of shapes for $H(z)$. Therefore, we anticipate that our results are general and generic for all similar parametrizations.

An interesting extension of our study would be to conduct an MCMC analysis considering the full CMB, the Pantheon and the growth data, as well as the BAO measurements. A similar analysis has been performed in \citep{Alestas:2020mvb} where it has been demonstrated that by considering an $H_0$ prior the best fit values of $(w_0, w_1)$ approach those described here, via our semi-analytic method. Alternatively, if one does not consider an $H_0$ prior the best fit values of $(w_0, w_1)$ end up close to the \plcdm model. This is due to the domination of the CMB - BAO data. Thus, such an MCMC exteneded analysis would include an $H_0$ prior along with the marginalization over other parameters demonstrating the growth tension level after such marginalization.

Our conclusion appears to favor a recently proposed generically distinct approach to the Hubble tension based on a rapid transition of SnIa absolute luminosity at $z\simeq 0.01$ due to a rapid change of the value of the gravitational constant by about $10\%$. This class of models has the following advantages over both early time and late time deformations of $H(z)$: It fully resolves the Hubble tension while also addressing the growth tension \citep{Marra:2021fvf, Alestas:2020zol}, it provides equally good fit to low z data (BAO and SnIa) as the \plcdm model, it has very interesting theoretical implications with respect to fundamental physics and finally it is testable by upcoming data and especially standard sirens data.

\section*{Acknowledgements}

We thank Savvas Nesseris and Lavrentios Kazantzidis for their useful input during the MCMC analysis, as well as Eoin \'O. Colg\'ain for useful discussions. All the MCMC chains were produced in the Hydra cluster of the Institute of Theoretical Physics (IFT) in Madrid, using \href{https://github.com/brinckmann/montepython_public}{MontePython}/\href{https://github.com/lesgourg/class_public}{CLASS} \citep{Brinckmann:2018cvx, Audren:2012wb, Blas_2011}. This research is co-financed by Greece and the European Union (European Social Fund - ESF) through the Operational Programme "Human Resources Development, Education and Lifelong Learning 2014-2020" in the context of the project  MIS 5047648.

\section*{Data Availability}

Access to the Pantheon compilation of SnIa data is provided in the \href{https://github.com/dscolnic/Pantheon}{Pantheon} Github repository, with their description being found \href{https://archive.stsci.edu/prepds/ps1cosmo/index.html}{here} and in Ref. \cite{Scolnic:2017caz}. The BAO data that were used in the analysis can be found in Refs. \citep{Anderson:2013zyy, Agathe:2019vsu, Percival_2010, 10.1111/j.1365-2966.2011.19250.x, Kazin:2014qga, Ross:2014qpa, Zhao:2016das, Bautista:2017zgn, Ata:2017dya, Abbott:2017wcz, Bautista:2017wwp, Gil-Marin:2018cgo, Zhao:2018gvb}. The CMB shift parameters used, are part of the Planck 2018 data release \citep{Aghanim:2018eyx}. Lastly, the numerical data files for the reproduction of the figures can be found in the \href{https://github.com/GeorgeAlestas/Growth-Hubble_Tension}{Growth-Hubble\_Tension} Github repository under the MIT license.



\bibliographystyle{mnras}
\bibliography{Bibliography} 



\bsp	
\label{lastpage}
\end{document}